\documentclass[conference]{IEEEtran}
\IEEEoverridecommandlockouts
\usepackage{enumitem} 
\usepackage[ruled,linesnumbered]{algorithm2e}
\usepackage{cite}
\usepackage{amsmath,amssymb,amsfonts}
\usepackage{algorithmic}

\usepackage{graphicx}
\usepackage{textcomp}
\usepackage{xcolor}
\usepackage{multirow}

\def\BibTeX{{\rm B\kern-.05em{\sc i\kern-.025em b}\kern-.08em
    T\kern-.1667em\lower.7ex\hbox{E}\kern-.125emX}}
\begin{document}
{
\title{Boosting Code-Switching ASR with Mixture of Experts Enhanced Speech-Conditioned LLM
\\
}


\author{\IEEEauthorblockN{Fengrun Zhang, Wang Geng\textsuperscript{*}, Hukai Huang, Yahui Shan, Cheng Yi, He Qu \thanks{*Corresponding author.}}
\IEEEauthorblockA{
\textit{Kuaishou Technology, Beijing, China} \\ 
\{gengwang\}@kuaishou.com
}
}
}

\maketitle

\begin{abstract}
In this paper, we introduce a speech-conditioned Large Language Model~(LLM) integrated with a Mixture of Experts (MoE) based connector to address the challenge of Code-Switching~(CS) in Automatic Speech Recognition~(ASR).
Specifically, we propose an Insertion and Deletion of Interruption Token~(IDIT) mechanism for better transfer text generation ability of LLM to speech recognition task. We also present a connecter with MoE architecture that manages multiple languages efficiently. To further enhance the collaboration of multiple experts and leverage the understanding capabilities of LLM, we propose a two-stage progressive training strategy: 1\text{)} The connector is unfrozen and trained with language-specialized experts to map speech representations to the text space. 2\text{)} The connector and LLM LoRA adaptor are trained with the proposed IDIT mechanism and all experts are activated to learn general representations. Experimental results demonstrate that our method significantly outperforms state-of-the-art models, including end-to-end and large-scale audio-language models. 

\end{abstract}


\section{Introduction}
Recently advancements in Speech Conditioned Large Language Models~(SC-LLM) have marked a significant milestone for Automatic Speech Recognition~(ASR)~\cite{wu2023decoder,tang2024salmonn,bai2024seed,geng2024unveiling}. Conditioned on speech representation and instruction, Typical SC-LLMs autoregressively generate speech transcriptions as the answer. 
Previous end-to-end models~\cite{dong2018speech,gulati20_interspeech,yao2023zipformer} are limited by the scarcity of labeled speech data due to the from-scratch training manner. Pre-trained on massive text data, LLMs process powerful capabilities for natural language processing~(NLP). 
Benefiting from LLMs' ability to understand and generate human-like text, SC-LLM emerges as a promising framework for next-generation ASR, which may be a potential breakthrough for some challenges, such as code-switching~(CS), multilingual and contextual biasing~\cite{yang2024mala} ASR.


CS scenarios refer to situations where a speaker switches between two or more languages within a single sentence. 
Although recent years have seen more and more effort dedicated to addressing this issue, 
nowadays CS scenarios remain challenging due to the following difficulties:

\begin{enumerate}[label=(\arabic*), leftmargin=0.5cm]
\item \textbf{Phonemic confusion.} A speaker might switch between languages that share acoustic similities, which makes it challenging to correctly detect the language and recognize the words. Mixture of Experts~(MoE) based models, as referenced in~\cite{lu2020bi,chen2023ba,ye2024sc}, have been proposed to alleviate this issue.
\item \textbf{Insufficient data.} Related studies were limited by insufficient data, particularly in scenarios where specific languages are interspersed, such as when Mandarin is mixed with English. 
\item \textbf{Model design omits CS scenarios.}  Nowadays, large-scale speech models, such as Qwen-Audio~\cite{chu2023qwen} and Whisper~\cite{whisper}, due to the label design of language identification, also perform poorly in CS scenarios.
\end{enumerate}
Since LLMs are pre-trained on multilingual text, including a substantial amount of code-switched text, they inherently possess the capability to understand and generate text that covers multiple languages. We consider this characteristic beneficial for further mitigating the aforementioned issues.

While previous works have demonstrated the powerful potential of SC-LLM in general ASR tasks~\cite{bai2024seed,geng2024unveiling}, their performance on CS scenarios remains largely uncharted. In this paper, we pioneer the exploration of their potential in Madarin-English CS scenarios. Inspired by the success of the MoE architecture in E2E models, we propose an SC-LLM integrated with a MoE-based connecter to enhance the model's capabilities. We also propose the Insertion and Deletion of Interruption Token (IDIT) mechanism, which better transfers the LLMs' text generation capability to ASR via constraining on predicted tokens. Furthermore, we introduce a two-stage progressive training strategy: 
an alignment stage to map the speech representation to the text space with language-specialized experts~(LSE); and a finetuning stage to enhance the collaboration among experts and achieve higher accuracy with IDIT.
Our contributions are summarized as follows:
\begin{itemize}
    \item We introduce an MoE-integrated SC-LLM to achieve high recognition accuracy in code-switching scenarios. 
    \item We propose an IDIT mechanism to effectively transfer the LLMs' text understanding ability to speech recognition.
    \item We propose a two-stage progressive training strategy to facilitate better modules collaboration. Our ablation study has demonstrated that this method enhances the model's capabilities in code-switching scenarios. 
    \item Our method significantly surpasses other models with a relative improvement of over 10\% on ASRU-2019 Mandarin-English
code-switching dataset, which demonstrates the potential of SC-LLM.
\end{itemize}


\section{Proposed Method}
\subsection{Overview}
\begin{figure}[t]
  \centering
  \includegraphics[width=\linewidth]{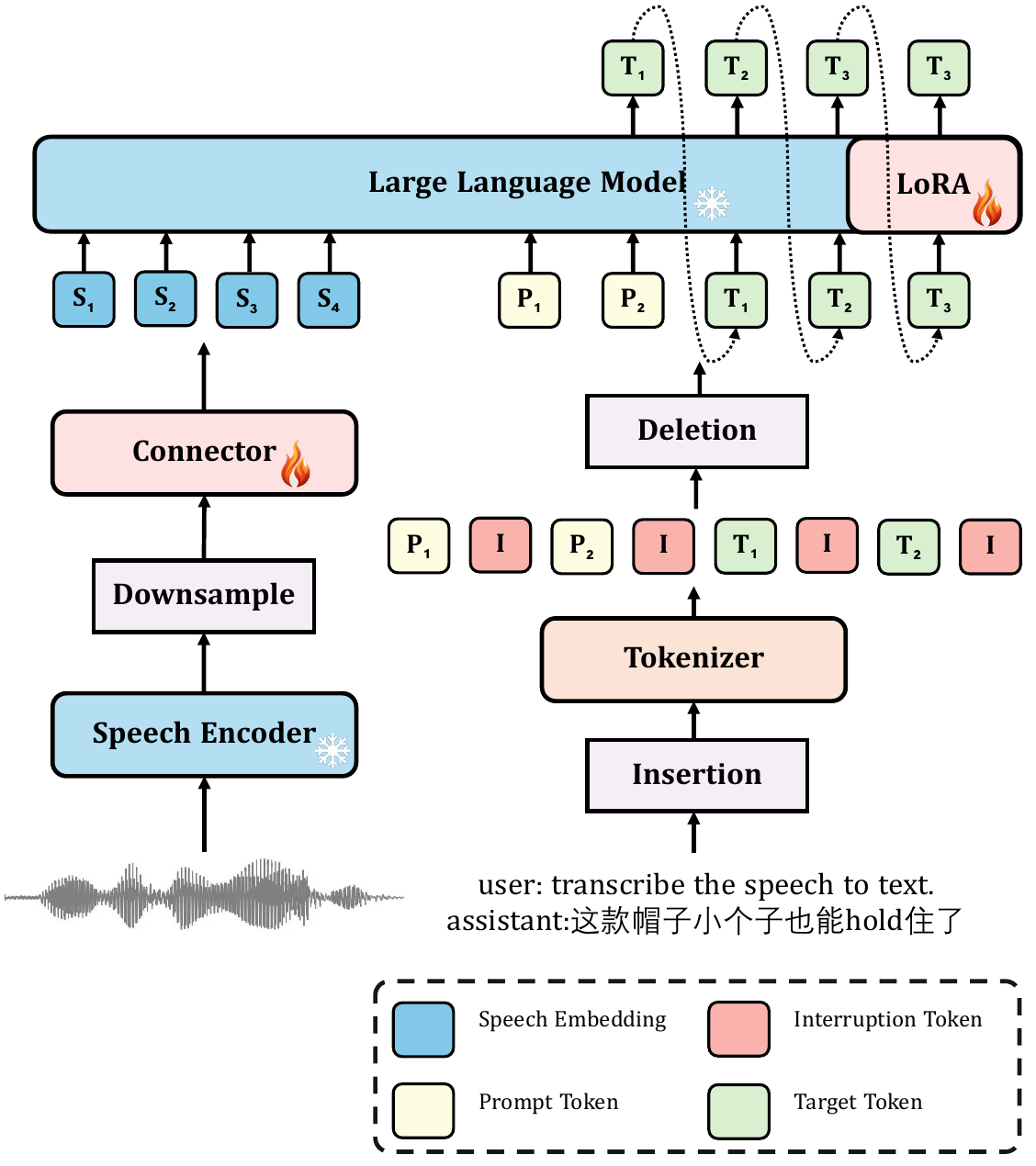}
  \caption{Overall Framework.} 
  \label{fig1}
\end{figure}

\vspace{-0.2cm}
An overview of our proposed model is shown in Fig.~\ref{fig1}. In our framework, speech encoders~\cite{hsu2021hubert, whisper,chen2022wavlm} are utilized to extract robust speech representations. Subsequently, the connector is designed to map downsampled speech representations into the text space, thereby aligning the different modalities and facilitating integration. Then the speech representation and instruction are concatenated to form a prompt and fed into a transformer-based LLM~\cite{vaswani2017attention}. The corresponding transcription is first converted into token IDs using the proposed IDIT mechanism and the tokenizer of LLM. Finally, these token IDs are fed into LLM as labels and generated via the next-token prediction process. With the learnable parameters from the connector and the Low-Rank Adaptation (LoRA)~\cite{hu2022lora} adaptor, the framework is trained to predict tokens autoregressively.

\subsection{Insertion and Deletion of Interruption Token}
Subword tokenization has been widely adopted in contemporary LLMs, including techniques such as WordPiece, Unigram and Byte-Pair Encoding~(BPE) tokenization~\cite{zhao2023survey}. These tokenizers aim to achieve better text compression, thereby improving the efficiency of both training and inference processes. Moreover, tokenization ensures the preservation of semantic integrity. The vocabulary size of these tokenizers is typically substantial, for example, Qwen2-7B~\cite{yang2024qwen2} has a vocabulary with 151,643 tokens. 

By experience, we have identified several key characteristics of these tokenizers: 1) due to their large vocabulary size, frequently used words are encoded into a single tokens examplified by the Chinese word 'nihao'. 2) they differentiate in encoding English words based on case sensitivity.
 Since words with uppercase letters occur less frequently in the English corpus, they are typically split into word pieces and encoded as multiple tokens.
 For instance, while Qwen2-7B~\cite{yang2024qwen2} tokenizer encodes 'speech' as a single token,  it encodes 'SPEECH' into three separate tokens: 'S', 'PE', and 'ECH'. 
 
 We consider the aforementioned characteristics to be inconsistent with speech recognition. In human speech perception, it is typical to segment Chinese at the character level and English at the word level. This distinction is also incorporated with the CS-ASR evaluation metrics, where Character Error Rate (CER) is used for Chinese and Word Error Rate (WER) for English. Therefore we believe that improving the tokenization process will better align the tokenizer with the speech recognition task.

Hence, we introduce the IDIT mechanism, which tokenizes Chinese at the character level and English at the word level. The details of this approach are elaborated in the algorithm~\ref{alg1}.



        
        

\begin{algorithm}[h]
	\caption{IDIT approach}
	\label{alg1}
	\KwIn{Transcription $\boldsymbol{X}$, Interruption Token $ <|I|>$ with token id $ i$, Tokenizer}
	\KwOut{Encoded token ids $\boldsymbol{Y}$.}  
	\BlankLine
	
	Given formatted text $\boldsymbol{X}'$ = None;
	
	Add a special token $<|I|>$ for Tokenizer to serve as an interruption;
	
	\ForEach{En word or Ch character $x_i$ in $\boldsymbol{X}$}
	{ 
	    $\boldsymbol{X}'$ = $\boldsymbol{X}'$ + $x_i$ + $ <|I|>$;
	}
	
	$\boldsymbol{Y}'$ = Tokenizer($\boldsymbol{X}'$);
	
	$\boldsymbol{Y}$ = $\boldsymbol{Y}'$.remove($i$);
	
	\Return $\boldsymbol{Y}$;
	
\end{algorithm}

The design of interruption token $ <|I|>$ allows the tokenizer to encode text precisely at the predefined character or word level. 
This design clarifies the task for LLM, which involves the autoregressive generation of Chinese characters or English words. 
This mechanism mitigates the challenge of learning the actual text length corresponding to each token, thereby reducing insertion and deletion errors.

 \subsection{MoE-based Connector}
An MoE-based connector is proposed to map the speech representations to multilingual embeddings, specifically tailored for code-switching scenarios. The connector is based on~\cite{tang2022sparse}, which comprises a weight-shared router and multiple experts.  The input is fed into the router, which predicts the probability of assigning each frame to a specific expert. Subsequently, the outputs from the activated experts are combined using the weighted probabilities. This process can be formulated as follows:
\begin{equation}
    \label{eq1}
    \mathcal{P}(x_i)=\mathrm{softmax}(\mathbf{W}x_i + b)
\end{equation}

\begin{equation}
    \label{eq2}
    h_i=\sum_{j=1}^n \mathcal{P}(x_i)_j \cdot e_j(x_i)
\end{equation}

where $ x_i\in\mathbf{R}^{d\times 1}$ is $i^{th}$ frame from speech representation $ X\in\mathbf{R}^{d\times t}$. The parameters $ \bold{W}\in\mathbf{R}^{n\times d}$ and $ b\in\mathbf{R}^{n\times 1}$ project the information from speech representation into the probability of selected experts $e_{1:n}$.

Moreover, We design a two-step method to enhance the code-switching representations. The two steps are illustrated in Fig.~\ref{fig3} and detailed as follows.
Specifically, We categorize parallel experts as language-specific experts to make better use of the monolingual and CS data in the training set. Experts are categorized into language-specialized $e_{ch}$ for Mandarin and $ e_{en}$ for English. 

In step 1, the process for the monolingual input can be formulated as follows:
\begin{equation}
\label{eq3}
h_i=\begin{cases}{ e_{ch}(x_i),  x_i \in Chinese}\\ 
{e_{en}(x_i) , x_i \in English  }
\end{cases}
\end{equation}
Meanwhile, frames from CS samples are routed into a selected expert with the maximum probability and concatenated along the original temporal dimension as the output $ H_{cs}$.

In step 2,
all experts are activated and the output is obtained via Equation (\ref{eq1}) and (\ref{eq2}). While strengthened in their respective languages, the experts then focus on learning generalized representations across multilingual inputs, thereby enhancing collaboration and generalization.

\subsection{Two-stage Progressive Training Strategy}
To harness the potential of multiple experts in CS scenarios, we propose a two-stage progressive training strategy to enhance collaboration and generalization among the experts. 
The first stage maps speech representation to text representation with language-specialized experts. The second stage integrates the IDIT mechanism and LoRA adapter to better align the output to speech recognition.

\begin{figure}[t]
  \centering
  \includegraphics[width=\linewidth]{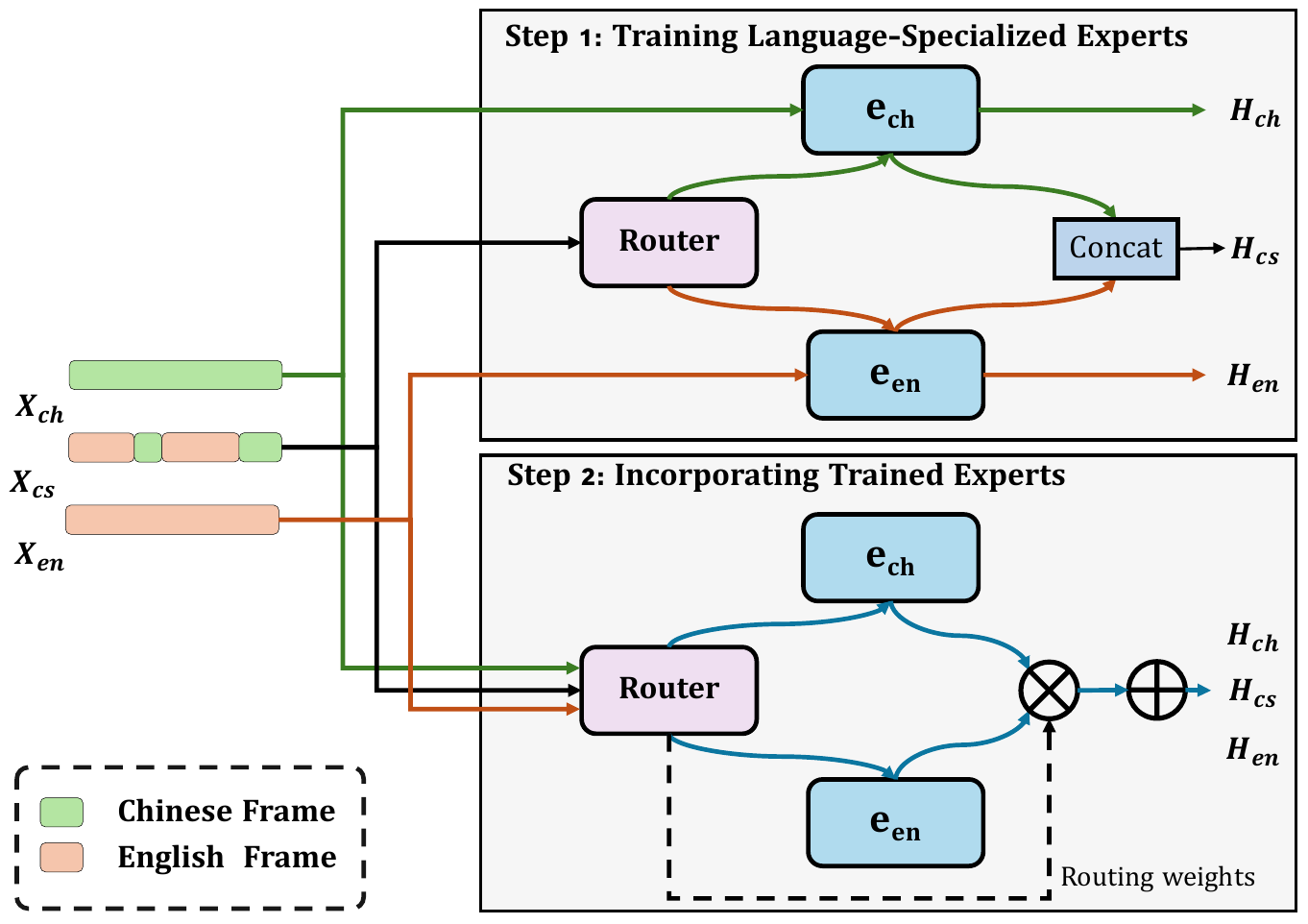}
  \caption{Two-step training for the proposed connector. In step 1, each frame is fed to only one expert. In step 2, each frame is fed to all experts, and the final output is derived from the summed outputs of multiple experts. } 
  \label{fig3}
\end{figure}

In the first stage, we aim to refine each expert’s proficiency by LSE with step 1 in Fig.~\ref{fig3}. 
Additionally, we do not use the IDIT mechanism to encode transcriptions at this stage. This means that the tokenizer encodes transcriptions consistent with the massive pre-trained data. This process aims to ensure that the connector is better trained with the semantic understanding provided by LLM.

In the second stage, the connector is trained with step 2 in Fig.~\ref{fig3} to learn generalized representations. Besides, we apply the IDIT mechanism to constrain the predicted tokens to word level for English or character level for Chinese. To bridge the gap between the LLM's original text generation capabilities and the predicted granularity better suited for speech recognition, we employ LoRA fine-tuning further to adapt the LLM to predict speech transcriptions. 
Finally, the predicted granularity is clarified via the IDIT mechanism to further enhance the capabilities.


\section{Experimental Setups}
\subsection{Datasets}
We conduct experiments on the ASRU-2019 Mandarin-English code-switching dataset~\cite{shi2020asru}, which comprises approximately 200 hours of code-switching training data and an additional 500 hours of Mandarin monolingual training data. Consistent with previous work~\cite{ye2024sc,chen2023ba}, we further add 460 hours of English monolingual data from Librispeech. The test set comprises 20 hours of code-switching data, containing approximately 16,000 utterances.

\subsection{Configuration}
All experiments in our work adhere to the following configurations unless otherwise specified. We adopt Whisper-large V3\footnote{https://github.com/openai/whisper} as the speech encoder and Qwen2-7B\footnote{https://github.com/QwenLM/Qwen2} as the LLM. The connector consists of two experts, and the output dimension of the router is equal to the number of experts. Each expert is a Feedforward Neural Network~(FFN) with an intermediate dimension configured to 2048. The downsampling module operates through a factor of 5 via frame splicing. For the LoRA configuration, we set $\alpha = 32$ and $ r = 8$. We set the batch size of 6 with a gradient accumulation of 3, resulting in a final batch size of 18. We use AdamW optimizer with the following settings: $\beta=(0.9, 0.999)$, $lr=5e-5$, and zero weight decay. Additionally, We employ the warmup schedule for the first 1000 training steps. Our models are trained on eight A800 GPUs. For the proposed two stages, each stage is trained for 1 epoch, encompassing approximately 25,000 steps. Our experiments are primarily conducted on SLAM-LLM toolkit\footnote{https://github.com/X-LANCE/SLAM-LLM}.

For evaluation metrics in code-switching scenarios, we calculate CER for the Chinese part, WER for the English part, and mixture error rate~(MER) for mixed scenarios.

\section{Experimental Results}

\subsection{Main Results}
We compare our method with recently prominent models in Table~\ref{tab1}, including E2E models and large-scale multilingual audio models. Note that the predictions from large-scale models are normalized as the final outputs.
\vspace{-0.2cm}
\begin{table}[ht]
\centering
\caption{Performance comparison. * denote that evaluations are performed ourselves. "Trainable" refers to the total number of trainable parameters in the model.}
\begin{tabular}{lc | ccc}
\label{tab1}
\\ \hline 
Model  & Trainable & CER(\%) & WER(\%) & MER(\%) \\ \hline
Conformer\textsuperscript{*}    &72M & 7.16           &     28.92     & 9.52  \\
SC-MoE     & 83.8M         & 6.50       &  \textbf{26.36}   &  8.66    \\
Whisper-large V3\textsuperscript{*} & 1.55B & 6.28   & 36.62        & 9.59    \\
Qwen2-Audio\textsuperscript{*}  & 8.2B   & \textbf{4.80}   & 55.23    & 10.31   \\ 
SenseVoice-small\textsuperscript{*} & 234M & 4.82   & 40.45   & 8.69        \\
\hline
\textbf{Ours} &\textbf{43.4M} & 5.13    & 29.36   & \textbf{7.76}  \\ \hline
\end{tabular}
\end{table}

As shown in Table~\ref{tab1}, our method achieves SOTA performance with fewer trainable parameters. 
In Table~\ref{tab1}, Conformer~\cite{gulati20_interspeech} is recognized as one of the most popular E2E models for general ASR. SC-MoE~\cite{ye24_interspeech} represents the SOTA E2E model specifically tailored for code-switching scenarios.
Models such as Whisper-large V3~\cite{whisper}, Qwen2-audio~\cite{qwen2-audio}, and SenseVoice-small~\cite{sensevoice} are capable of performing multiple speech-related tasks. Even though these tasks are typically beneficial for multilingual alignment and understanding, the scarcity of CS data and the design for language identification tasks render these models susceptible to language confusion in CS scenarios. 
Our method offers a more effective perspective by leveraging robust speech representations, cross-modal alignment, and LLMs' powerful understanding abilities.
\subsection{Connector Architectures}
To verify the effectiveness of the proposed MoE-based connector, we compare it with the widely used linear layer in Table~\ref{tab2} which consists of a single FFN layer. Furthermore, we introduced the Aishell-1 test~\cite{bu2017aishell} for Mandarin and Librispeech test-clean for English~\cite{panayotov2015librispeech} to evaluate the monolingual capabilities. 
As can be seen in Table~\ref{tab2}, both monolingual and code-switching abilities are boosted with the integration of MoE. 
 \vspace{-0.2cm}
\begin{table}[ht]
\centering
\caption{Performance comparison over different connectors.}
\begin{tabular}{lcc|cc}
\label{tab2}
\\ \hline 
Dataset & Language & Metric & Linear  & MoE\\ \hline
Aishell-1  & CH   & CER~(\%)  & 5.60         &  \textbf{5.42}   \\
Librispeech      & EN  & WER~(\%)      & 2.79        &  \textbf{2.68} \\ 
ASRU     & CH-EN   & MER~(\%) & 8.10        &  \textbf{7.76} \\ \hline
\end{tabular}
\end{table}

\subsection{Ablation Study on Training Strategies}
We evaluate the effectiveness of our proposed method by showcasing various strategies in Table~\ref{tab3}. 
As strategy $F$ shows the proposed two-stage progressive training strategy achieves the best performance, we believe that employing LSE to initialize the connector and employing IDIT for the final predicted granularity is more effective.
As shown in Table~\ref{tab3}, enabling only 
IDIT (strategy $B$) or LSE (strategy $C$) throughout the entire process
is not optimal, which results in performance
 degradation compared with strategy a.
 Instead, enabling only LSE in stage1 (strategy $D$) or IDIT in stage2 (strategy $E$)
obtains higher performance compared with strategy $A$.
Moreover, strategy $F$ leads to the best performance, suggesting that the initial application of LSE to enhance 
the connector's ability, followed by the subsequent use of the IDIT mechanism
to predict speech-level tokens, is highly effective within the training process.

 \vspace{-0.2cm}
\begin{table}[ht]
\centering
\caption{Ablation study on training strategies.}
\begin{tabular}{c c c c c | c}
\label{tab3}
\\ \hline 
\multirow{2}{*}{Strategy} & \multicolumn{2}{c }{Stage1} & \multicolumn{2}{c | }{Stage2} & \multirow{2}{*}{MER~(\%)}\\ 
& IDIT& LSE & IDIT & LSE &  \\ \hline
$A$  & $\times $ & $\times $ & $\times $   & $\times $&   8.47   \\
$B$  &  $\checkmark $ & $\times $ &  $\checkmark $   & $\times $&   9.26   \\ 
$C$ &  $\times $&    $\checkmark $ &   $\times $ &  $\checkmark $  & 8.51 \\ 
$D $&  $\times $ &     $ \checkmark $  & $\times$  & $\times$ & 8.31\\ 
$E$ &  $\times $ &    $\times$  & $ \checkmark $  &   $\times$ & 7.94\\ 
$F $&  $\times $&  $\checkmark $ &   $\checkmark $  & $\times $&  \textbf{7.76}  \\  \hline
\end{tabular}
\end{table}

\subsection{Generalizability}
To assess the generalizability of our proposed method, we conduct experiments by replacing the speech encoder with Hubert\footnote{https://github.com/MoBofeng/chinese-hubert-large} or LLM with Baichuan2-7B\footnote{https://github.com/baichuan-inc/Baichuan2}. As shown in Table.~\ref{tab4}, our method delivers competitive results when integrated with various encoders or LLMs.
 \vspace{-0.2cm}
\begin{table}[ht]
\centering
\caption{Performance with different modules.}
\begin{tabular}{lc | c}
\label{tab4}
\\ \hline 
Encoder      & LLM  &    MER~(\%)  \\  \hline
Hubert &  Qwen2 & 8.69 \\
Whisper  & Baichuan2    &   8.33 \\
Whisper     & Qwen2   & \textbf{7.76} \\  \hline
\end{tabular}
\end{table}

\section{Conclusion}
In this paper, we introduce a speech-conditioned LLM enhanced by MoE to address the challenge of Code-Switching Speech Recognition. The innovation of our proposed method lies in the design of the interruption token and the two-stage progressive training with a MoE-based connector. The interruption token forces the LLMs' tokenizer to encode the transcriptions with word level for English and character level for Chinese, thereby boosting the recognition accuracy. Additionally, the code-switching ability is further developed by the MoE-based connector with language-specified experts. Experimental results on Mandarin-English code-switching datasets demonstrate superior performance over existing models.

\bibliographystyle{IEEEtran}
\bibliography{mybib}

\end{document}